# Fog-assisted Caching Employing Solar Renewable Energy for Delivering Video on Demand Service

Mohamad Bin Abdull Halim[1], Sanaa Hamid Mohamed[2], Taisir E. H. El-Gorashi[2], and Jaafar M. H. Elmirghani[2]

[1] *Device Development Group, Intel Microelectronics, Penang, Malaysia*
[2] *School of Electronic and Electrical Engineering, University of Leeds, LS2 9JT, United Kingdom*

**ABSTRACT**

This paper examines the reduction in brown power consumption of transport networks and data centres achieved by caching Video-on-Demand (VoD) contents in solar-powered fog data centers with Energy Storage Devices (ESDs). A Mixed Integer Linear Programming (MILP) model was utilized to optimize the delivery from cloud or fog data centres. The results reveal that for brown-powered cloud and fog data centres with same Power Usage Effectiveness (PUE), a saving by up to 77% in transport network power consumption can be achieved by delivering VoD demands from fog data centres. With fully renewable-powered cloud data centres and partially solar-powered fog data centres, savings of up to 26% can be achieved when considering 250 $m^2$ solar cells. Additional saving by up to 14% can be achieved with ESDs of 50 kWh capacity.

## 1. INTRODUCTION

The video traffic is estimated to have a Compound Annual Growth Rate (CAGR) of 54% from 2016 to 2021 [1]. As a result, the power consumption of transport networks linking cloud data centres and end users is expected to massively increase. As these systems are typically brown-powered, this also leads to steep rise in $CO_2$ emission and operational costs due to high utilization and cooling requirements against thermal dissipation [2]. To overcome both issues, several greening approaches were considered in the last decade such as improving the hardware, optimizing the routing and workload scheduling, and using renewable power sources [3]. The authors of [2] considered lightpath bypassing in IP over WDM core networks to reduce the power consumption of the non-bypass approach. As part of the outcomes of GreenTouch, a leading Information and Communication Technology (ICT) research consortium with 50 industrial and academic collaborators, the work in [4]-[16] investigated a combination of greening approaches for IP over WDM networks. Those included optical bypassing, topology optimizations, Mixed Line Rates (MLRs), efficient protection and sleep modes, in addition to considering two improvement schemes for hardware which are the Business-As-Usual (BAU) improvement in equipment due to CMOS technology advances, and BAU with further GreenTouch improvements. The former indicated 4.23× energy efficiency improvements compared to 2010 networks while the later indicated 20× improvements.

Optimizing the workloads and content placement to reduce the traffic and hence the power consumption was also considered to green core networks as in [17]–[19]. In [17], the authors focused on data centre and popular contents placement strategies and found that placing the data centres at the centre of the network and replicating the contents on multiple data centres according to popularity minimized the power consumption by 28%. In [18] and [19], the caching of Video-on- Demand (VoD) contents is optimized to reduce storage and transport energy consumption while considering sizes of the caches, contents popularity at different hours and dynamic cache contents replacement. To reduce the $CO_2$ emission coupled with the rise in brown power consumption, the use of renewable resources was considered to power different networking and data centre elements [20]-[23].

Different implementations such as Mobile Edge Computing (MEC), Fog Computing (FC), and cloudlet Computing (CC) were recently introduced to reduce the latency of cloud computing [24] and improve the energy efficiency of transport networks [25]-[28]. Nano Data Centres (NaDa) were introduced in [29] as a Peer-to-peer (P2P) computing and storage infrastructure and energy consumption reductions by 20-30% were obtained. Using fog data centres for smart cities was discussed in [30] to reduce core networks power consumption. The performance and power consumption tradeoffs of using different data centre topologies for big data computations in fog environments was discussed in [31]. In [32], the concept of integrating micro data centre (Micro-DC) into Optical Line Terminals (OLTs) of Passive Optical Networks (PONs) was discussed to partially reduce core networks traffic. To enhance the use of interrupted renewable sources such as solar power for data centres, optimizing the use of Energy Storage Devices (ESDs) was suggested [33].

This work utilizes a MILP model to reduce the brown power consumption of transport networks when delivering VoD contents by maximizing the use of solar energy in fog data centres with ESDs in the access network. The rest of this paper is organized as the following: Section 2 elaborates on the system model and the parameters used in the MILP model. The results are presented in Section 3, while, the conclusions and future work are given in Section 4.

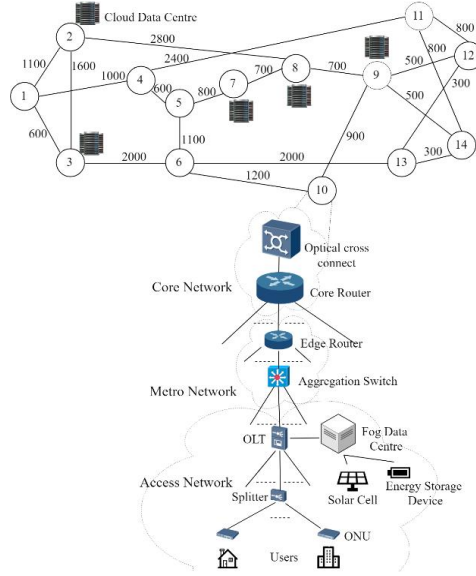

*Figure 1: Fog data centre caching model to assist cloud VoD services.*

## 2. SYSTEM MODEL FOR OPTIMIZING VOD DELIVERY FROM CLOUD OR FOG DATA CENTRES

An IP over WDM network with NSFNET topology was utilized for the core network as shown in Figure 1. Core nodes are equipped with MLR IP router and transponder ports and the links have adequate EDFAs and regenerators. All devices have power consumption values taken from [5] for 2020 equipment. Cloud data centers (CDCs) are pre-located in nodes 2, 3, 7, 8, and 9. In each core node, a metro network, composed of edge routers and Ethernet switches (C9500-32QC [34]), is utilized to provide connection to the access networks. The access network is composed of OLTs [35] connecting the metro network with Fog Data Centres (FDCs), in addition to splitters and ONUs connecting to end users. For CDCs and FDCs, networking equipment power consumption is assumed to be 30% of the servers' power [36]. The content server in [23] which has a maximum capacity of 1.8 Gbps was considered, which allows FDCs to maximally provide 160 Gbps via about 88 servers. We considered solar renewable energy for its suitability in fog environments within cities. The solar irradiance values in all 14 nodes were collected from [37] and an efficiency of 26.3% was considered [38]. Each FDC is powered by brown sources, and directly by solar cells with areas between 50 and 250 $m^2$, or additionally by stored solar energy in an ESD (e.g. Li-ion battery) with capacities between 20 and 50 kWh [39]. Power Usage Effectiveness (PUE) values between 1.25 and 1.1 for FDCs and of 1.1 for CDCs were considered. Consumer video traffic based on Cisco Visual Network Index (VNI) forecast [5] was considered for the demands.

*Table 1. Key Parameters for the MILP Model.*

| | |
|---|---|
| Power consumption of a metro Ethernet switch 40 Gbps port [34] | 50 W |
| Power consumption of a content server per Gbps [23] | 221.1 W |
| Capacity of a content server [23] | 1.8 Gbps |
| PUE of cloud data centre ($PUE_c$) | 1.1 |
| PUE of fog data centre ($PUE_f$) | 1.25 to 1.1 |
| Ratio to account for networking equipment power consumption [36] | 1.3 |
| Power consumption of an OLT [35] | 904 W |
| Total capacity of links between OLT and metro network | 160 Gbps |
| Total capacity of links between OLT and fog data center | 160 Gbps |
| Size of solar cell per OLT ($SSC$) | 50, 100, 150, 200, 250 $m^2$ |
| Battery maximum capacity ($E_{max}$) [39] | 20-50 kWh |
| Charging percentage per hour and Discharging percentage per hour [33] | 72.25%, and 90.25% |
| Self-discharging per day [33] | 3% |

## 3. RESULTS

*A. Power consumption with brown-powered data centres:*

We start by evaluating the brown power consumption ($PC_b$) required to optimally deliver VoD demands in terms of power consumption efficiency from brown-powered CDCs and FDCs. Figure 2 shows the total $PC_b$ per day for different $PUE_f$ values. The results show that for $PUE_f$ of 1.25, delivering fully from CDCs is the most efficient. As $PUE_f$ improve, the model starts to deliver partially from FDCs. When $PUE_f$ is equivalent to $PUE_c$, it becomes more efficient to fully stream from FDCs as the required power consumption to deliver from FDCs and CDCs will be equivalent, and the transport network consumption will be the factor determining the differences in $PC_b$.

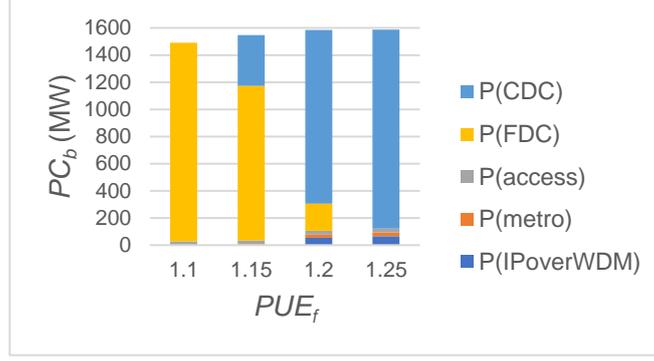

Figure 2: $PC_b$ under different $PUE_f$ values.

*B. Power consumption with fully renewable-powered CDCs and renewable-powered FDCs:*
We now consider fully renewable-powered CDCs and FDCs with $PUE_f$ of 1.1 and solar cells. Figure 3a shows the total $PC_b$ per day when considering different sizes for the solar cells (i.e. $SSC$). The results indicate that savings by up to 26% can be achieved in the transport network relative to case A when fully delivering from the CDCs.

*C. Power consumption with fully renewable-powered CDCs and caching in renewable-powered FDCs with ESDs:*
In this case we consider optimizing the streaming of VoD from cloud or fog data centers with $PUE_f$ of 1.1 and $SSC$ of 250 $m^2$ while optimizing ESDs usage. Figure 3b shows the total $PC_b$ per day when considering different capacities for Li-ion batteries. The results indicate that additional savings by up to 14% can be achieved in the transport network relative to case B for solar cell size of 250 $m^2$. The increase in power savings values is due to optimizing the direct use of solar power in the FDC or charging the ESD for the use when it is not available.

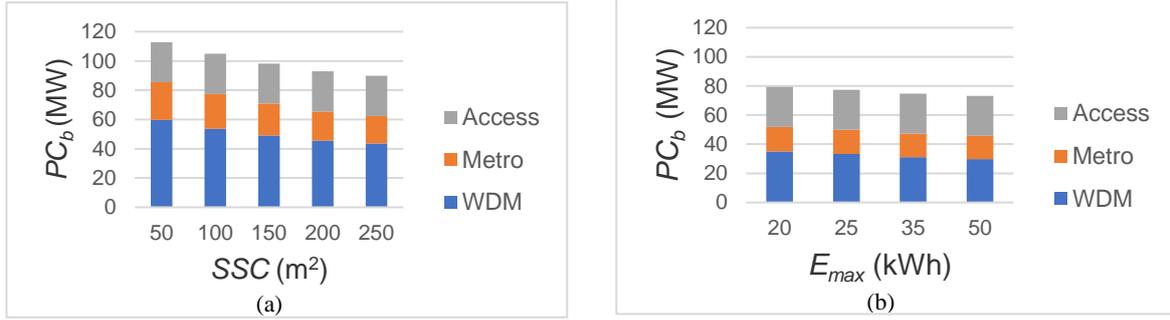

Figure 3: (a) $PC_b$ with renewable-powered CDCs and partially solar-powered FDCs with different SSC values, (b) $PC_b$ with renewable-powered CDCs and partially solar-powered FDCs with ESDs.

## 4. CONCLUSIONS AND FUTURE WORK

This paper addressed the optimization of delivering VoD services from cloud or fog data centres with solar cells and ESDs. With brown-powered data centres, the results indicate that with $PUE_f$ higher than $PUE_c$, it is more energy efficient to deliver partially or fully from CDCs. When $PUE_f$ is equivalent to $PUE_c$, it is more efficient to deliver from FDCs. As many cloud providers utilize renewable power, we also examined the optimization when CDCs are fully renewable powered and the FDCs are partially solar powered. Savings by up to 26% can be achieved when considering 250 $m^2$ solar cells and additional saving by up to 14% can be achieved when also considering ESDs with capacity of 50 kWh. Future work includes considering the actual networking power consumption of different topologies for FDCs, and the storage requirements and popularity of VoD contents.

**ACKNOWLEDGEMENTS**

The authors would like to acknowledge funding from the Engineering and Physical Sciences Research Council (EPSRC) for the INTERNET (EP/H040536/1), and STAR (EP/K016873/1). All data are provided in full in the results section of this paper.

**REFERENCES**

[1] "Cisco Visual Networking Index: Global Mobile Data Traffic Forecast Update, 2016-2021," White Paper, Cisco, March 2017.
[2] G. Shen and R. Tucker, "Energy-Minimized Design for IP Over WDM Networks," Optical Communications and Networking, IEEEIOSA Journal of, vol. 1, no. 1, pp. 176-186, June 2009.

[3] Y. Zhang, P. Chowdhury, M. Tornatore, and B. Mukherjee, "Energy Efficiency in Telecom Optical Networks," Communications Surveys Tutorials, IEEE, vol. 12, no. 4, pp. 441-458, Fourth 2010.
[4] J. M. H. Elmirghani, L. Nonde, A. Q. Lawey, T. E. H. El-Gorashi, M. O. I. Musa, X. Dong, K. Hinton, and T. Klein, "Energy efficiency measures for future core networks," in 2017 Optical Fiber Communications Conference and Exhibition (OFC), March 2017, pp. 1-3.
[5] J. M. H. Elmirghani, T. Klein, K. Hinton, L. Nonde, A. Q. Lawey, T. E. H. El-Gorashi, M. O. I. Musa, and X. Dong, "GreenTouch GreenMeter core network energy-efficiency improvement measures and optimization," IEEE/OSA Journal of Optical Communications and Networking, vol. 10, no. 2, pp. A250-A269, Feb 2018.
[6] M. Musa, T.E.H. El-Gorashi and J.M.H. Elmirghani, "Bounds on GreenTouch GreenMeter Network Energy Efficiency," *IEEE/OSA Journal of Lightwave Technology*, vol. 36, No. 23, pp. 5395-5405, 2018.
[7] H.M.M., Ali, A.Q. Lawey, T.E.H. El-Gorashi, and J.M.H. Elmirghani, "Future Energy Efficient Data Centers With Disaggregated Servers," *IEEE/OSA Journal of Lightwave Technology*, vol. 35, No. 24, pp. 5361 – 5380, 2017.
[8] X. Dong, T.E.H. El-Gorashi and J.M.H. Elmirghani, "On the Energy Efficiency of Physical Topology Design for IP over WDM Networks," *IEEE/OSA Journal of Lightwave Technology*, vol. 30, pp.1931-1942, 2012.
[9] B. Bathula, M. Alresheedi, and J.M.H. Elmirghani, "Energy efficient architectures for optical networks," Proc IEEE London Communications Symposium, London, Sept. 2009.
[10] M. Musa, T.E.H. El-Gorashi and J.M.H. Elmirghani, "Bounds for Energy-Efficient Survivable IP Over WDM Networks with Network Coding," *IEEE/OSA Journal of Optical Communications and Networking*, vol. 10, no. 5, pp. 471-481, 2018.
[11] B. Bathula, and J.M.H. Elmirghani, "Energy Efficient Optical Burst Switched (OBS) Networks," *IEEE GLOBECOM'09*, Honolulu, Hawaii, USA, November 30-December 04, 2009.
[12] X. Dong, T.E.H. El-Gorashi and J.M.H. Elmirghani, "Green Optical OFDM Networks," *IET Optoelectronics*, vol. 8, No. 3, pp. 137 – 148, 2014.
[13] M. Musa, T.E.H. El-Gorashi and J.M.H. Elmirghani, "Energy Efficient Survivable IP-Over-WDM Networks With Network Coding," IEEE/OSA Journal of Optical Communications and Networking, vol. 9, No. 3, pp. 207-217, 2017.
[14] A. Lawey, T.E.H. El-Gorashi, and J.M.H. Elmirghani, "BitTorrent Content Distribution in Optical Networks," *IEEE/OSA Journal of Lightwave Technology*, vol. 32, No. 21, pp. 3607 – 3623, 2014.
[15] A.M. Al-Salim, A. Lawey, T.E.H. El-Gorashi, and J.M.H. Elmirghani, "Energy Efficient Big Data Networks: Impact of Volume and Variety," IEEE Transactions on Network and Service Management, vol. 15, No. 1, pp. 458 - 474, 2018.
[16] A.M. Al-Salim, A. Lawey, T.E.H. El-Gorashi, and J.M.H. Elmirghani, "Greening big data networks: velocity impact," IET Optoelectronics, vol. 12, No. 3, pp. 126-135, 2018.
[17] X. Dong, T. El-Gorashi, and J. Elmirghani, "Green IP Over WDM Networks With Data Centers," Lightwave Technology, Journal of, vol. 29, no. 12, pp. 1861-1880, June 2011.
[18] N. I. Osman and T. El-Gorashi and J. M. H. Elmirghani, "The impact of content popularity distribution on energy efficient caching," in 2013 15th International Conference on Transparent Optical Networks (ICTON), June 2013, pp. 1-6.
[19] N. I. Osman, T. El-Gorashi, L. Krug, and J. M. H. Elmirghani, "Energy Efficient Future High-Definition TV," Journal of Lightwave Technology, vol. 32, no. 13, pp. 2364-2381, July 2014.
[20] X. Dong, T. El-Gorashi, and J. Elmirghani, "IP Over WDM Networks Employing Renewable Energy Sources," Lightwave Technology, Journal of, vol. 29, no. 1, pp. 3-14, Jan 2011.
[21] L. Nonde, T. E. H. El-Gorashi, and J. M. H. Elmirgahni, "Virtual Network Embedding Employing Renewable Energy Sources," in 2016 IEEE Global Communications Conference (GLOBECOM), Dec 2016, pp. 1-6.
[22] L. Nonde, T.E.H. El-Gorashi, and J.M.H. Elmirghani, "Energy Efficient Virtual Network Embedding for Cloud Networks," *IEEE/OSA Journal of Lightwave Technology*, vol. 33, No. 9, pp. 1828-1849, 2015.
[23] A. Q. Lawey, T. E. H. El-Gorashi, and J. M. H. Elmirghani, "Renewable energy in distributed energy efficient content delivery clouds," in 2015 IEEE International Conference on Communications (ICC), June 2015, pp. 128-134.
[24] A. C. Riekstin, B. B. Rodrigues, K. K. Nguyen, T. C. M. de Brito Carvalho, C. Meirosu, B. Stiller, and M. Cheriet, "A Survey on Metrics and Measurement Tools for Sustainable Distributed Cloud Networks," IEEE Communications Surveys Tutorials, vol. 20, no. 2, pp. 1244-1270, Second quarter 2018.
[25] F. Jalali, K. Hinton, R. Ayre, T. Alpcan, and R. S. Tucker, "Fog computing may help to save energy in cloud computing," IEEE Journal on Selected Areas in Communications, vol. 34, no. 5, pp. 1728-1739, May 2016.
[26] A.N. Al-Quzweeni, A. Lawey, T.E.H. El-Gorashi, and J.M.H. Elmirghani, "Optimized Energy Aware 5G Network Function Virtualization," *IEEE Access*, vol. 7, 2019.

[27] M.S. Hadi, A. Lawey, T.E.H. El-Gorashi, and J.M.H. Elmirghani, "Patient-Centric Cellular Networks Optimization using Big Data Analytics," *IEEE Access*, vol. 7, 2019.

[28] M.S. Hadi, A. Lawey, T.E.H. El-Gorashi, and J.M.H. Elmirghani, "Big Data Analytics for Wireless and Wired Network Design: A Survey, Elsevier Computer Networks, vol. 132, No. 2, pp. 180-199, 2018.

[29] V. Valancius, N. Laoutaris, L. Massoulie, C. Diot, and P. Rodriguez, "Greening the Internet with Nano Data Centers," in Proceedings of the 5th International Conference on Emerging Networking Experiments and Technologies, ser. CoNEXT '09. New York, NY, USA: ACM, 2009, pp. 37-48.

[30] S. Igder, S. Bhattacharya, and J. M. H. Elmirghani, "Energy Efficient Fog Servers for Internet of Things Information Piece Delivery (IoTIPD) in a Smart City Vehicular Environment," in 2016 10th International Conference on Next Generation Mobile Applications, Security and Technologies (NGMAST), Aug 2016, pp. 99-104.

[31] S. H. Mohamed, T. E. H. El-Gorashi, and J. M. H. Elmirghani, "Energy Efficiency of Server-Centric PON Data Center Architecture for Fog Computing," in 2018 20th International Conference on Transparent Optical Networks (ICTON), July 2018, pp. 1-4.

[32] B. Yang, Z. Zhang, K. Zhang, and W. Hu, "Integration of micro data center with optical line terminal in passive optical network," in 2016 21st OptoElectronics and Communications Conference (OECC) held jointly with 2016 International Conference on Photonics in Switching (PS), July 2016, pp. 1-3.

[33] C. Gu, K. Hu, Z. Li, Q. Yuan, H. Huang, and X. Jia, "Lowering Down the Cost for Green Cloud Data Centers by Using ESDs and Energy Trading," in 2016 IEEE Trustcom/BigDataSE/ISPA, Aug 2016, pp. 1508–1515.

[34] Cisco Catalyst 9500 series switches data sheet. (Cited on 2018, Apr). [Online]. Available:https://www.cisco.com/c/en/us/products/collateral/switches/catalyst-9500-series-switches/datasheet-c78-738978.html

[35] ZXA10 C300: The Industry's First Future-proof Optical Access platform. (Cited on 2018, Apr). [Online]. Available: http://www.zte.com.cn/global/products/access/xpon/PON-OLT/424194

[36] M. Dayarathna, Y. Wen, and R. Fan, "Data Center Energy Consumption Modeling: A Survey," IEEE Communications Surveys Tutorials, vol. 18, no. 1, pp. 732–794, Firstquarter 2016.

[37] NREL: MIDC/NREL Solar Radiation Research Laboratory (BMS). (Cited on 2018, Feb). [Online]. Available: https://midcdmz.nrel.gov/apps/go2url.pl?site=BMS

[38] K. Yoshikawa, H. Kawasaki, W. Yoshida, T. Irie, K. Konishi, K. Nakano, T. Uto, D. Adachi, M. Kanematsu, H. Uzu, and K. Yamamoto, "Silicon heterojunction solar cell with interdigitated back contacts for a photo- conversion efficiency over 26%," Nature Energy, vol. 2, p. 17032, 2017.

[39] H. Chen, T. N. Cong, W. Yang, C. Tan, Y. Li, and Y. Ding, "Progress in electrical energy storage system: A critical review," Progress in Natural Science, vol. 19, no. 3, pp. 291 – 312, 2009.